\documentclass[aps,twocolumn,showlabels,showrefs,amsmath,amssymb,pre,superscriptaddress, floatfix, colors]{revtex4}

\usepackage{graphicx}
\usepackage{dcolumn}
\usepackage{bm}
\usepackage{amssymb}
\usepackage{multirow}
\usepackage{color}
\usepackage{hyperref}

\newcommand{\1}{\begin{equation}}
\newcommand{\2}{\end{equation}}
\newcommand{\ea}{\begin{eqnarray}} 
\newcommand{\ee}{\end{eqnarray}}
\newcommand{\4}[2]{{\frac{#1}{#2}}}
 
\newcommand{\I}{{ {\rm i}  }} 
\newcommand{\de}{{\rm d}}
\newcommand{\Sum}[2]{{\sum\limits_{#1}^{#2}}}

\begin{document}

\title{The Rotating Vicsek Model: Pattern Formation and Enhanced Flocking in Chiral Active Matter}
\date{\today}


\author{Benno Liebchen}\email[]{Benno.Liebchen@staffmail.ed.ac.uk}\affiliation{SUPA, School of Physics and Astronomy, University of Edinburgh, Edinburgh EH9 3FD, United Kingdom}
\author{Demian Levis}\email[]{levis@ub.edu}\affiliation{Departament de F\'isica de la Mat\`eria Condensada, Universitat de Barcelona, Mart\'i i Franqu\`es 1, E08028 Barcelona, Spain}

\begin{abstract}
We generalize the Vicsek model to describe the collective behaviour of polar circle swimmers with local alignment interactions. 
While the phase transition leading to collective motion in 2D (flocking) occurs at the same interaction to noise ratio as for linear swimmers, as we show, 
circular motion enhances the polarization in the ordered phase (enhanced flocking) and induces secondary instabilities leading to structure formation. 
Slow rotations promote phase separation whereas fast rotations 
generate patterns consisting of phase synchronized microflocks with a controllable self-limited size. 
Our results defy the viewpoint that monofrequent rotations form a vapid extension of the Vicsek model and establish
a generic route to pattern formation in chiral active matter 
with possible applications to control coarsening and to design rotating microflocks.
\end{abstract}

\maketitle

Among the most remarkable features of active matter systems is their ability to spontaneously form self-sustained nonequilibrium structures, without requiring external driving. 
These active structures range from motility-induced phase separation of self-propelled particles into a dense and a dilute phase \cite{Tailleur2008, Cates2015} and clusters
of self-limited size \cite{Theurkauff2012,Palacci2013,Buttinoni2013,Levis2014,Liebchen2015} in isotropic active matter, 
to long range ordered flocks and travelling bands in 2D polar active matter 
\cite{Vicsek1995,Toner1995, Farrell2012, Caussin2014, Solon2015}.
Despite their phenomenological diversity most of these (and other) activity-induced structures
can be observed in a small class of archetypical minimal models allowing to explore their universality. 
For linear self-propelled particles which change their swimming direction only by diffusion (and alignment interactions), 
the Active Brownian Particle model and 
the Vicsek model have become standard models representing isotropic and polar active matter.

Besides such linear swimmers, there is now a strong interest in a new class of self-propelled particles which change their direction of motion autonomously.
This class of chiral active matter includes a variety of biological circle swimmers, such as \emph{E.coli} which swim circularly 
when close to walls and interfaces 
\cite{Berg1990, diLuzio2005, Lauga2006, DiLeonardo2011}, as well as  
sperm cells \cite{Friederich2012, Riedel2005}, and magnetotactic bacteria in rotating external fields \cite{Erglis2007,Cebers2011}.
Following the general principle that any deviation between the self-propulsion direction of the particle and its symmetry axis couples
its translational and rotational degrees of freedom, it has also been possible to design synthetic circle swimmers; examples being 
L-shaped self-phoretic swimmers \cite{Kummel2013,Hagen2014} and actuated colloids allowing
to design radius and frequency of circular trajectories on demand.  
While these synthetic examples have supported the recent boost of interest in chiral active matter, as the recent reviews \cite{Loewen2016, Friederich2016} reflect, 
surprisingly little is known about their collective
behaviour (exceptions exploring collective behaviour are \cite{Denk2016,Liebchen2016}).

Therefore, following the spirit of formulating minimal models for the collective behaviour of linear active matter, we introduce here 
the '\emph{rotating Vicsek model}' (RVM) to 
describe the collective behaviour of polar circle swimmers. 
This model describes overdamped self-propelled particles changing their direction autonomously with an intrinsic rotation frequency,
and with local alignment interactions between circle swimmers (which are typically non-spherical). 
\\In the monofrequent case of identical circle swimmers, 
one might expect that circular swimming has little impact on the physics of the standard Vicsek model
as the absence of inertia seems to guarantee invariance of the system by global rotation of the reference frame -- as for an overdamped ideal gas in a rotating bucket, where global rotations 
do not change the particle dynamics inside. 
This viewpoint receives further support
by the fact that the flocking transition of the Vicsek models proves invariant under rotations, as we will show.  
Strikingly, however, this flocking transition induces long-range polar order, which spontaneously breaks 
rotational invariance and allows rotations to dramatically change the physics of the Vicsek model. 
When rotations are fast compared to rotational diffusion, which is a natural parameter range for many circle swimmers,
a new phase occurs, which we call the \emph{rotating micro-flock phase}.  
This phase emerges via a short-wavelength clustering instability from a uniform flock and leads to a proper pattern of localized rotating flocks 
which do not coarsen beyond a characteristic length scale. This scale increases linearly with the swimming speed and decreases with the rotational frequency, allowing to use 
rotations as a tool to design microflock patterns. 
Besides fast rotations, also slow ones induce interesting collective effects:
they allow for phase separation and lead to coherently moving large-scale structures with droplet-like shapes 
featuring an enhanced polarization as compared to flocks in the standard Vicsek model.
\\Thus, in contrast to the common opinion that identical circle swimmers do not change the collective behaviour of linear swimmers significantly, 
the present work shows that they lead to a rich new phase diagram, involving a novel route to pattern formation.
This route should be readily observable in identical synthetic circle swimmers (L-shaped or actuated colloids)  
or in magnetotactic bacteria in rotating external magnetic fields, and could be useful, for example, 
to design localized micro-flocks whose characteristic size can be (dynamically) controlled in the laboratory (e.g. by changing the self-propulsion velocity or the frequency of the applied field). 

Besides this, our results may find further applications for understanding pattern formation in 2D suspensions of sperm cells \cite{Riedel2005}
and driven protein filaments \cite{Loose2013, Denk2016} qualitatively matching the microflocks we observe.
In this context, we note that our results may qualitatively apply even to nonidentical but synchronized biological swimmer ensembles as discussed in \cite{Liebchen2016}.

\begin{figure}
\begin{center}
\includegraphics[scale=.43,angle=0]{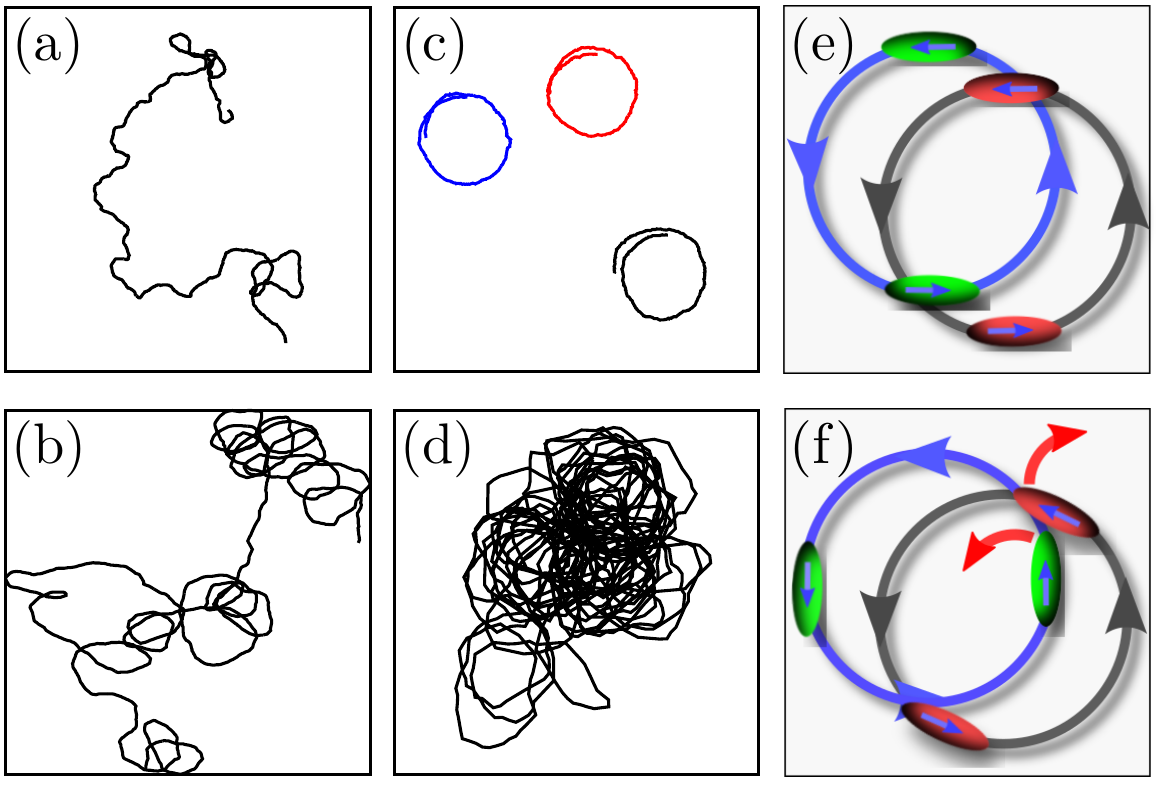}
\caption{\small {Trajectories of a linear (a, $\Omega=0$) and a circle swimmer (b, $\Omega=3$). 
(c): For slow rotations ($g=0.14$, $\Omega=0.2$), circle swimmers phase-lock and follow circular orbits allowing for 
aligned configurations (e) and the formation of large rotating droplets.
(d): Fast rotations ($g=0.14$, $\Omega=3$) leave no time to phase lock, which frustrates the alignment interactions and destroys circular trajectories (f).
Self-organizing into a microflock pattern where circle swimmers move irregularly around a common microflock-centers allows them to
compromise between rotations and alignment.}}
\label{cartoon}
\end{center}
\end{figure}

\begin{figure*}
\begin{center}
\includegraphics[scale=0.5,angle=0]{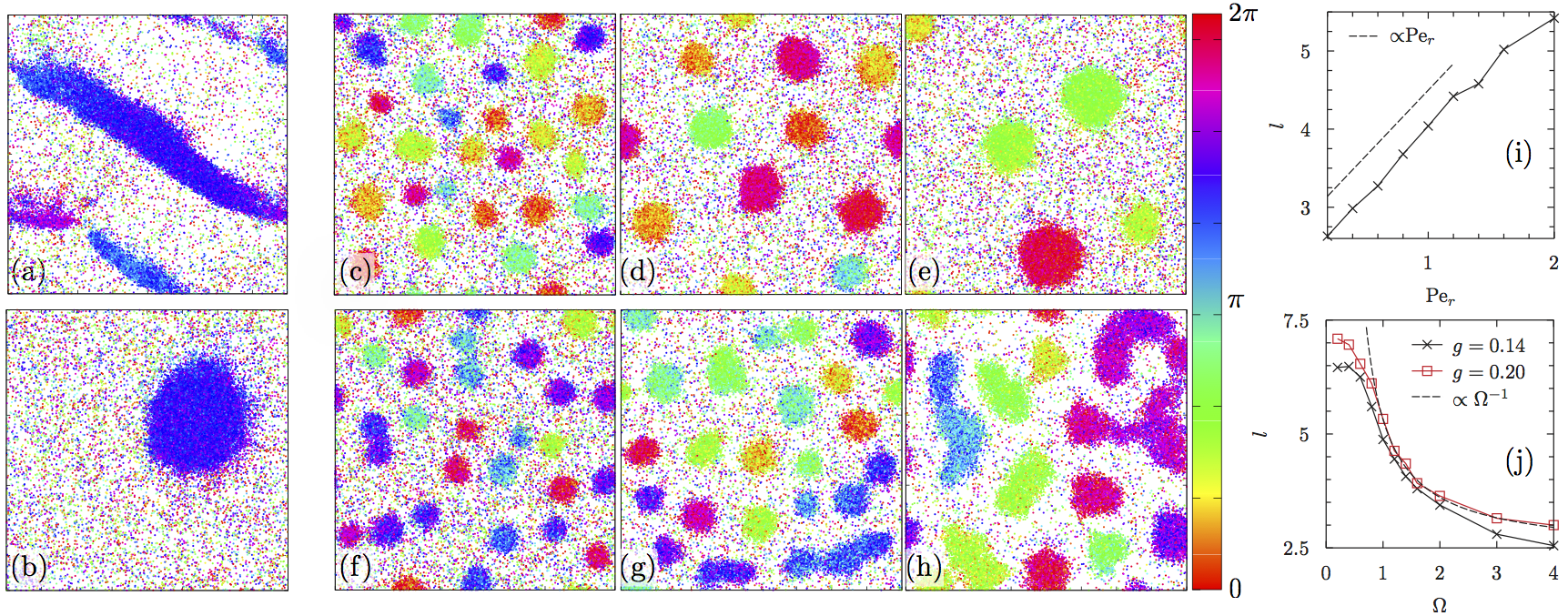}
\end{center}
\caption{\small
Simulation snapshots for $N=32000$ particles; colours encode particle orientations.
(a,$\Omega=0$): Travelling bands; (b, $\Omega=0.2<1$): rotating droplet (phase-separation) 
(c-h): Microflock pattern at $g=0.14, \Omega=3$ and Pe$_r=0.2$ (c), 
Pe$_r=1.0$ (d) and  Pe$_r=2$ (e) and at Pe$_r=0.2, \Omega=3$ and $g=0.12$ (f), $0.18$ (g) and $0.3$ (h). 
(i,j): Microflock length scale $l$ for $\Omega=3; g=0.14$ as a function of ${\rm Pe}_r$ (i) and for ${\rm Pe}_r=0.2$ as a function of $\Omega$ (j).} 
\label{fig:pattern}
\end{figure*}

\paragraph*{The rotating Vicsek model} 
To specify our results we now define the RVM: it consists 
of $N$ point-like self-propelled particles with positions ${\bf r}_i$ and orientations ${\bf p}_i(t)=(\cos \theta_i,\,\sin\theta_i)$
which interact via an aligning pair-potential and change their direction in response to a systematic rotational force, according to: 
\begin{equation}
\dot{\bf r}_i=v {\bf p}_i \, , \dot{{\theta}}_i=\omega +\frac{K}{\pi R_{\theta}^2}\sum_{j \in \partial_i}\sin(\theta_j-\theta_i)+\sqrt{2D_r}{\eta}_i\,,
\end{equation}
Here, the sum runs over neighbours within a radius $R_\theta$ around particle $i$ and ${\eta}_i(t)$ is a unit-variance Gaussian white noise with zero mean. 
In the non-interacting limit ($K=0$), each particle performs 
an overdamped circular Brownian motion as shown in Fig.~\ref{cartoon} and statistically characterised in \cite{vanTeeffelen2008}.
To reduce the parameter space to its essential dimensions, we 
choose space and time units as $R_\theta$ and $1/D_r$. 
The RVM has four control parameters: the particle density $\rho_0=N/L^2$, 
a Peclet number ${\rm Pe}_{r}=v/(D_r R_\theta)$ measuring the persistence length in units of the alignment interaction range,
$g=K/(\pi R_\theta^2 D_r)$ and $\Omega=\omega/ D_r$, comparing alignment and rotational frequencies with the rotational diffusion rate.  
Remarkably, the phase diagram depends only on $g\rho_0$ and $\Omega$, as we discuss below, with most interesting phenomena occurring for 
$g_f:=g\rho_0 > 2$ and for $\Omega \sim 1$ or $\Omega >1$.
While the former criterion is the flocking criterion of the standard Vicsek model, 
most circle swimmers naturally feature suitable $\Omega$ values:
Rotating $E.coli$ ($\omega\sim 0.1-1 /s$ \cite{Lauga2006}; $D_r \sim 0.2/s-1/s$) lead to 
$\Omega \sim 1$, whereas L-shaped swimmers ($\omega \sim 0.1-0.3/s$; $D_r\sim 6 .10^{-4}$ \cite{Kummel2013}) allow to explore the regime $\Omega \sim 10^2\gg 1$
and magnetotactic bacteria in rotating fields should allow to tune $\Omega$ on demand. 

\begin{figure}
\begin{center}
\includegraphics[width=0.48\textwidth]{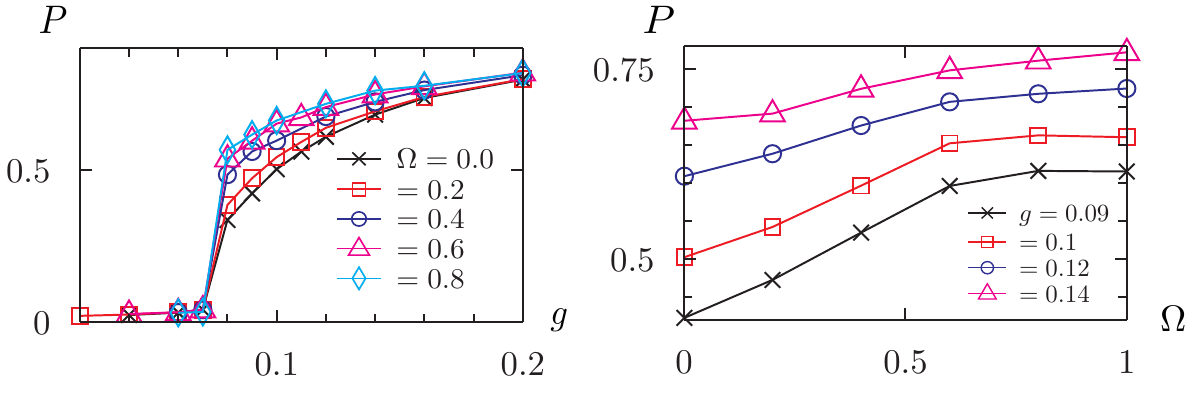}
\end{center}
\caption{\small 
Global polarization over $g$ (left) and $\Omega$ showing invariance of the flocking transition against rotations (left) and enhanced flocking (right) as predicted in the text.}
\label{fig:flocking}
\end{figure}

\paragraph*{Pattern formation}
We now simulate the collective behaviour of $N=32000$ identical circle swimmers in a quadratic box with periodic boundary conditions.
For $\Omega=0$ we reproduce the phenomenology of the standard Vicsek model \cite{VicsekRev, Chate2004, Mishra2010, Solon2015}: a disordered homogeneous phase occurs below the flocking threshold ($g<g_f$), 
whereas $g \gtrsim g_f$ induces a global polarization with high density bands coexisting with a disordered gas (Fig. \ref{fig:pattern} (a)). 
These bands eventually become unstable at higher coupling strengths, leading to homogeneous flocking. 
Now choosing $g>g_f$ and switching on slow rotations ($\Omega=0.2$),
we observe phase separation into a large polarly ordered dense phase and a low-density gas of incoherently rotating swimmers. 
Here, the presence of rotations changes the geometry of the high density region which now takes the form of a spherical cluster (droplet), reminiscent of the usual liquid-gas demixing.
This droplet rotates coherently but slower than individual swimmers with a frequency $\Omega^\ast< \Omega$ (see Fig. \ref{cartoon} (c), \ref{fig:pattern} (b) and Movie 1 in the Supplementary Material (SM) \cite{SM}). 
Tuning the frequency to values $\Omega \gtrsim 1$ arrests phase separation and leads, strikingly, to a pattern of dense spots which do not 
grow beyond a self-limited size (see Fig.~\ref{fig:pattern} (c)-(h) and Movie 2). 
Within each spot, particles are synchronized and form rotating microflocks: hence we call the emerging phase \emph{the rotating microflock pattern}.
This pattern resembles vortex arrays observed in sperm cells and protein filaments \cite{Riedel2005, Loose2013}. 
\paragraph*{Hydrodynamic equations and enhanced flocking}
To understand the emergence of patterns and their length scales, we derive a continuum theory for the RVM in the SM \cite{SM}. 
Following the approaches in \cite{Dean1996,Bertin2009}
we find \cite{SM} a closed set of equations for the particle density $\rho({\bf x},t)$ and polarization density ${\bf w}({\bf x},t)=(w_x,w_y)=\rho {\bf P}$ (with $ {\bf P}({\bf x},t)$ being the polarization field)
where $|{\bf w}|$ measures the local degree of alignment and ${\bf w}/|{\bf w}|$ the average swimming direction. 
\ea
\dot \rho &=& -{\rm Pe}_{\rm r} \nabla \cdot {\bf w} \label{rho1}\\
\dot {\bf w}& =& \left(g \rho-2\right)\4{{\bf w}}{2} - \4{{\rm Pe}_{\rm r}}{2}\nabla \rho + \4{{\rm Pe}^2_{\rm r}}{2 b} \nabla^2 {\bf w} - \4{g^2}{b}|{\bf w}|^2{\bf w}  \label{w1mt}\\
&+& \4{g {\rm Pe}_{\rm r}}{4 b}\left[5\nabla {\bf w}^2 - 10{\bf w}(\nabla \cdot {\bf w}) -6 ({\bf w}\cdot \nabla){\bf w}\right] \nonumber \\
&+&\Omega {\bf w}_\perp  +\4{{\rm Pe}_{\rm r}^2 \Omega}{4b}\nabla^2 {\bf w}_\perp - \4{g^2 \Omega}{2 b}|{\bf w}|^2 {\bf w}_\perp \nonumber \\ 
&-& \4{g {\rm Pe}_{\rm r} \Omega}{8 b}\left[3\nabla_\perp {\bf w}^2 - 6{\bf w} (\nabla_\perp \cdot {\bf w}) - 10({\bf w} \cdot \nabla_\perp) {\bf w} \right] \nonumber 
\ee
Here $b=2(4+\Omega^2)$, ${\bf w}^{(1)}_\perp=(-w_y^{(1)},w_x^{(1)})$ and $\nabla_\perp=(-\partial_y,\partial_x)$.
We first note that the disordered uniform phase (D)
$(\rho,{\bf w})=(\rho_0,{\bf 0})$ solves eq. (\ref{w1mt}) with $\rho_0$ being the particle density. 
Linearizing eq. (\ref{w1mt}) around D (SM \cite{SM}) unveils an instability (flocking transition)
$g\rho_0>2$, which is the same as for linear swimmers ($\Omega=0$) showing that the emergence of long-range order is invariant to rotations. 
Our simulations confirm this invariance (Fig.~\ref{fig:flocking}).\footnote{We find a flocking transition close to but slightly below 
the theoretical prediction, as previously noted in \cite{Farrell2012}.}. 
Following the flocking instability, the RVM approaches a rotating uniform phase (F), $(\rho,|{\bf w}|,{\bf w}/|{\bf w}|)=(\rho_0,w_0,\cos(\Omega_0 t),\sin(\Omega_0 t))$, 
featuring long-range order:
\1
w_0 = \4{1}{g}\sqrt{\left(g\rho_0 - 2\right) \left(4+\Omega^2 \right)} \label{fpol}
\2
In this phase, a macroscopic fraction of circle swimmers phase-synchronizes and rotates coherently with a frequency 
$\Omega_0 = \Omega \left(\4{3}{2}-\4{g \rho_0}{4}\right)$.
This frequency reduces to the single particle frequency at the onset of flocking, but slows
down as $g\rho_0$ increases.
Remarkably, Eq.~(\ref{fpol}) suggests that the polarization increases with $\Omega$, a phenomenon which we call \emph{enhanced flocking} and confirm using particle based simulations in Fig.~\ref{fig:flocking}.
\footnote{Note that the system typically does not reach F but forms secondary structures due to instabilities of F. 
However, enhanced polarization can still be observed for the (locally uniform) bubbles.}
Physically, enhanced flocking might be based on a decrease of the average time needed for a diffusive rotating particle (which is not yet part of the flock) to align its direction with the flock. 
That is, rotations allow the flock to collect particles with random orientations faster.  

\paragraph*{Microflock-instability}
To understand the transition from (F) to the patterns observed above, we now perform a linear stability analysis of F. 
Here, the presence of long-range order allows terms of order $\Omega {\bf w}\nabla_\perp {\bf w}$ to crucially impact the stability of (F) as we will see.
First considering the case $\Omega=0$ we find an oscillatory 
long wavelength instability along the polarization direction for $2<g \rho_0 < 22/7$ 
(and a stationary long wavelength instability perpendicular to the flocking direction for $2<g\rho_0<82/21$).
The oscillatory instability evokes moving density fluctuations only in polarization direction and is often associated with the 
emergence of travelling bands in the standard Vicsek model \cite{Bertin2009,Mishra2010}.
In the RVM we also find oscillatory long wavelength instabilities, here producing moving density fluctuations both longitudinal and perpendicular to the flocking direction
allowing for (coarsening) rotating droplets 
(Fig.~{\ref{fig:pattern}})b in the RVM.

Most strikingly, for larger $\Omega$ our  
linear stability analysis (\cite{SM}) unveils a rotation-induced oscillatory short wavelength 
instability. 
This instability generates pattern formation in the RVM and explains the observation of microflocks with 
a self-limited size (Fig. (\ref{fig:pattern})); hence we call it the microflock-instability.
Close to $g\rho_0=2$ the characteristic microflock size scales as (see \cite{SM})
\1 
l \approx \4{\pi{\rm Pe}_r}{2\Omega^2} \4{|4(2-g\rho_0)+\Omega^2(12-g\rho_0)|}{\sqrt{(g\rho_0-2)(4+\Omega^2)}} \label{mfscl}
\2
Thus, microflocks grow linearly with ${\rm Pe}_r$, but also grow with $g\rho_0$ and decrease with $\Omega$ in most parameter
regimes. If $\Omega\gg 1$, (\ref{mfscl}) yields 
$l^\ast \propto v/\omega$: i.e. for fast rotations, 
the microflock size is proportional to the radius of a single circle swimmer.
Our simulations confirm all these scalings (Fig.~\ref{fig:pattern} (i-j)): 
Specifically, defining the length scale $l$ of a numerically observed structure 
as the value of $l$ where the pair correlation function $G(l)=1$ leads to 
Fig.~\ref{fig:pattern}: panel (i) confirms the $l \propto {\rm Pe}_r$ prediction and (j) shows a decrease of $l$ 
with increasing $\Omega$, revealing that the microflock size 
can be tuned by the microscopic parameters in our model.\footnote{In Fig.~\ref{fig:pattern} (j) 
we only show $l$ within the regime where microflocks are approximately isotropic. For larger $g$, the length scale $l$ as defined by the pair 
correlation function depends on the microflock shape and doesn't represent their length scale in a unique way.}
Note, that the microflock-instability does not only provide a proper route to pattern formation but also allows for 
structure formation at interaction to noise ratios where the standard Vicsek model is deep in the uniform flocking phase. 

What is the physical mechanism leading to the rotating droplet phase and the microflock pattern?
While circle swimmers are effectively independent of each other at large distances in phase (D), 
for $g\rho_0>2$ they have satisfy the rotations while being aligned on average. 
If interactions dominate ($g\rho_0/\Omega \gg 1$) circle swimmers can phase lock before they rotate much and follow almost ideal circles (Fig.~\ref{cartoon} (c)). Here, 
they are parallel to each other all along their circular orbits (Fig.~\ref{cartoon} (e)) and form a macroscopic rotating droplet (Fig.~\ref{fig:pattern}(b)).
In this state, interactions support circular motion: phase locking leads to an essentially stiffly rotating many-particle object that experiences an 'average' noise,
inducing only weak deviations from circular motion (Fig.~\ref{cartoon} (c)).
Conversely, when rotations dominate ($g\rho_0/\Omega < 1$), 
the phase locking timescale
becomes comparable to the rotational timescale. 
This results in phase shifts among adjacent circle swimmers that frustrate, for swimmers on circular orbits, the alignment interaction (Fig.~\ref{cartoon} (f)). 
The frustration, in turn, destroys circular orbits and makes large droplets of phase-locked swimmers impossible.
As a result, the droplet phase breaks down which opens a route to pattern formation: the resulting microflock phase can be seen as an 
attempt of the RVM to satisfy alignment interactions in presence of rotations but in absence of phase-locking, at least on average (see Fig.~\ref{cartoon} (d) for a typical trajectory):
rotating around a common center allows particles to avoid close-to-orthogonal configurations as the one shown in Fig.~\ref{cartoon} (f) even in 
presence of small phase shifts. 
Increasing the size of a microflock therefore dissatisfies the alignment interactions; hence microflocks naturally resist coarsening beyond a certain scale.

\begin{figure}
\begin{center}
\includegraphics[width=0.4\textwidth]{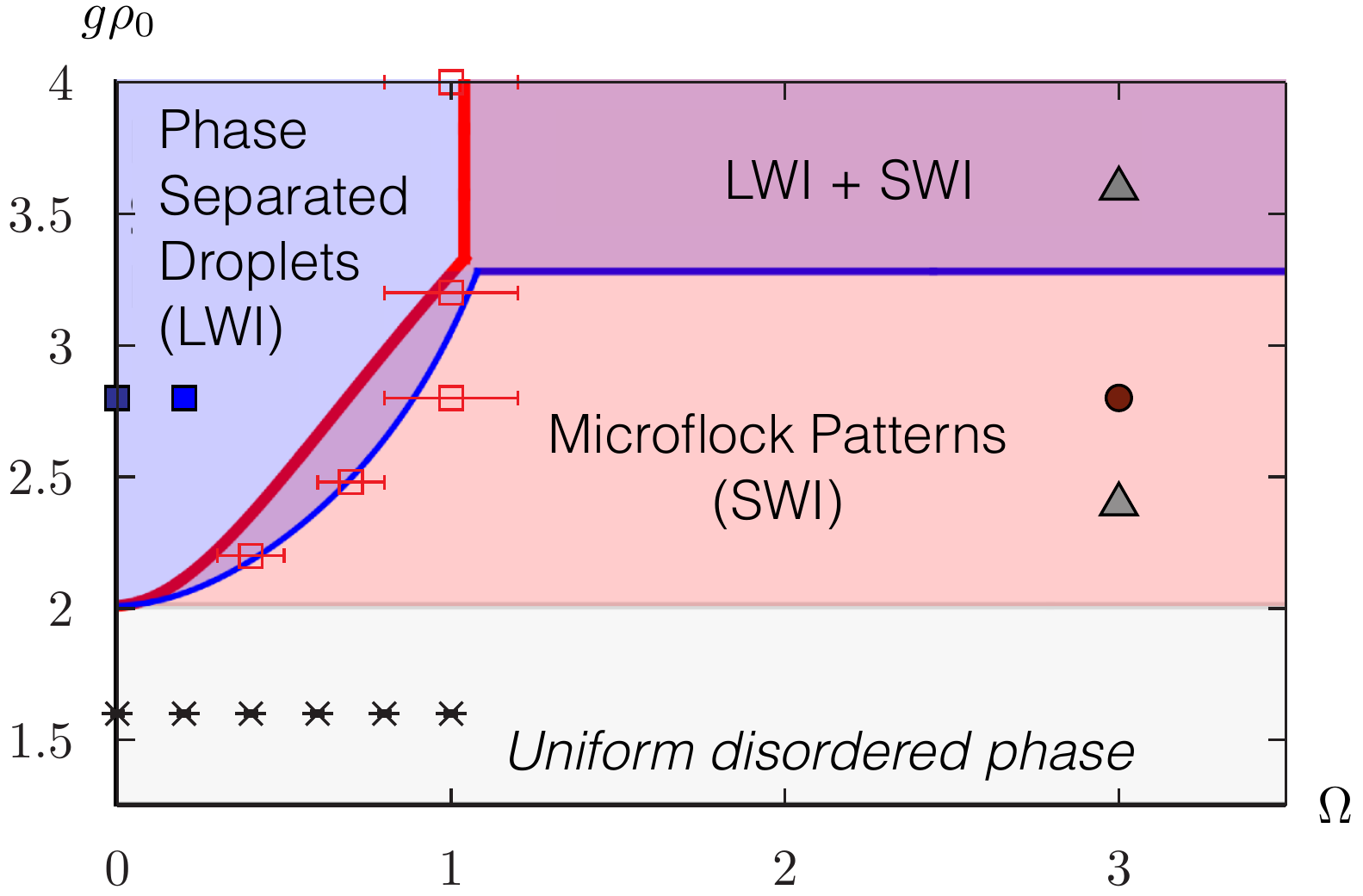}
\end{center}
\label{fig4}
\caption{\small Nonequilibrium phase diagram.
Red domain: Oscillatory, short wavelength instability (SWI) inducing microflock patterns; blue region: phase-separating droplets induced by long wavelength instabilities 
(LWI; perpendicular to flocking direction in \cite{SM}). 
Red symbols show simulation results for the microflock-droplet-transition. 
Grey domain: stability of uniform disordered phase; black crosses: flocking transition from simulations. 
Filled symbols show parameters of Fig. \ref{fig:pattern}: (a,b) blue squares; (c-e) brown dot, (f-g) grey triangles.
} 
\label{fig:phd}
\end{figure}

To get an overview of the parameter regimes leading to droplet and microflock patterns 
we summarize our results from linear stability analysis and 
simulations in an instability or phase diagram, Fig.~\ref{fig:phd}. 
Although the RVM depends on four dimensionless parameters, we show in the SM \cite{SM} that its phase diagram 
is fully characterized by $g\rho_0$ and $\Omega$. Thus, the two-dimensional plot in Fig.~\ref{fig:phd} represents the whole parameter space. 
In this plot, red shaded areas lead to pattern formation while blue ones represent the rotating macrodroplet phase (phase separation). 
Where both regimes overlap ($\Omega \sim 1$ and $g\rho_0 \gtrsim 10/3$) short and long wavelength instabilities perpendicular to the flocking 
direction coexist. Generally, we also find a coexisting long wavelength instability in polarization direction, 
which is not shown in Fig.~\ref{fig:phd} but detailed in the SM \cite{SM}.
Often, the coexisting long and short wavelength instabilities 
are separated by a band of stable wavenumbers (Fig.~1 in \cite{SM}), suggesting 
that, depending on initial conditions, (F) proceeds either to phase separation or to pattern formation. 
This suggests hysteresis in the RVM: we confirm this in Movie 3, showing 
phase separation for small $\Omega$ persisting even after a quench to large $\Omega$ values, which 
normally lead to the microflock pattern, when our system is initialized in phase (F).

\paragraph*{Conclusions} 
Conversely to the viewpoint
that identical rotations are unimportant for the collective behaviour of overdamped self-propelled particles, 
we show they generate a generic route to structure formation. 
While slow rotations promote phase separation yielding a 
rotating macrodroplet featuring an enhanced polarization compared to the standard Viczek model, faster 
rotations induce phase-synchronized microflocks with a self-limited size.
This size can be 
tuned via the swimming speed and the rotation frequency allowing to use rotations as a design principle for microflock patterns. 
Our results should be observable, e.g. with autophoretic L-shaped colloids or magnetotactic bacteria, and
provide a general framework to acknowledge and understand the rich collective behaviour of chiral active matter. 

\paragraph*{Acknowledgements}
BL and DL gratefully acknowledge funding from a Marie Curie Intra European Fellowship (G.A. no 654908 and G.A. no 657517) within Horizon 2020.
\\BL and DL contributed equally to this work.


%

\newpage
\onecolumngrid

\begin{center}
{\large \bf Supplementary Material
\vspace{0.4cm}\\The Rotating Vicsek Model: Pattern Formation and Enhanced Polarization in Chiral Active Matter}
\end{center}



\section{Continuum Theory of Circle Swimmers}
Here, we develop a continuum theory for the rotating Vicsek model (RVM), closely following the approach in \cite{Liebchen2016sm}.
We start with the Langevin equations as given by Eqs. (1) in the main text
but
replace the 
finite range alignment interaction by a pseudopotential ('$\delta$'-interaction), which is justified if the 
interaction is short ranged enough, 
such that the shape of the associated interaction potential is irrelevant to the many particle dynamics. Using dimensionless units, this leads to the following Langevin equations
\1
\dot {\bf r}_i = {\rm Pe}_{\rm r} {\bf p}_i; \quad \dot \theta_i = \Omega + g\Sum{j\neq i}{}\delta({\bf r}_j-{\bf r}_i)\sin(\theta_j-\theta_i) + \sqrt{2}\xi_i(t)\ .
\2
where $\xi_i(t)$ represents Gaussian white noise with zero mean and unit variance. 

Now using It\^{o}s Lemma and following \cite{Dean1996sm}
we derive a continuum equation of motion for the combined $N$-particle probability density 
$f({\bf r},\theta,t)=\Sum{i=1}{N}\delta({\bf r}-{\bf r}_i(t))\delta(\theta-\theta_i(t))$ of finding a circle swimmer with 
orientation ${\bf p}=(\cos \theta, \sin \theta)$ at position ${\bf r}$ at time $t$:
\ea
\dot f = - {\rm Pe}_{r}{\bf p}\cdot \nabla f - \Omega \partial_\theta f - g \partial_\theta \int \de \theta' f({\bf r},\theta') \sin(\theta'-\theta) f({\bf r},\theta) + \partial^2_\theta f - \partial_\theta \sqrt{2 f} \eta
\label{fkfp}
\ee
Here $\eta= \eta({\bf r},\theta,t)$ is a unit-variance Gaussian white noise field with zero mean. 
In the following, we focus on a mean-field description and neglect the multiplicative noise term $-\partial_\theta \sqrt{2f}\eta$.
Transforming
(\ref{fkfp}) to Fourier space, yields an equation of motion for the Fourier modes $f_k({\bf r},t)=\int f({\bf r},\theta,t){\rm e}^{\I k \theta} \de \theta$
of $f$:
\1
\dot f_k({\bf r},\theta,t) =-\4{{\rm Pe}_{\rm r}}{2} \left[\partial_x \left(f_{k+1}+f_{k-1}\right) - \I \partial_y \left(f_{k+1}-f_{k-1}\right)\right]+(\I k \Omega f_k - k^2) f_k + 
\4{\I g k}{2\pi} \Sum{m=-\infty}{\infty}f_{k-m} F_{-m} f_m
\label{fkfpk}
\2
Here $F_m$ is the $m$-th Fourier coefficient of $\sin(\theta)$. 
Evaluating (\ref{fkfpk}) for $k=0,1..$ leads to a hierarchy of equations for $\{f_k\}$ with $f_0({\bf x},t)=\rho({\bf x},t)=\int f({\bf x},\theta,t)\de \theta$ 
being the probability density to find a circle swimmer at time $t$ at position ${\bf x}$
(independently of its orientation) 
and $({\rm Re}f_1,{\rm Im}f_1)={\bf w}({\bf x},t)=\int {\bf p}(\theta) f({\bf x},\theta,t) \de \theta$ is the polarization density: the magnitude $w\equiv |{\bf w}|$ 
represents the fraction of aligned circle swimmers and ${\bf w}/w$ their average swimming direction. 
To close the hierarchy of equations (\ref{fkfpk}) we follow the scheme of \cite{Bertin2009sm}, involving the assumption that 
deviations from isotropy are not too strong. Specifically, we assume that $f_2$, representing nematic order, follows changes in $f_0,f_1$ adiabatically (i.e. $\dot f_2\approx 0$) and that 
higher order fields approximately vanish ($f_{k_\geq 3} \approx 0$). 
After a long but straightforward calculation, we find the following equations of motion for $\rho, {\bf w}$
\ea
\dot \rho &=& -{\rm Pe}_{\rm r} \nabla \cdot {\bf w} \label{rho1}\\
\dot {\bf w}& =& \left(g \rho-2\right)\4{{\bf w}}{2} - \4{{\rm Pe}_{\rm r}}{2}\nabla \rho + \4{{\rm Pe}^2_{\rm r}}{2 b} \nabla^2 {\bf w} - \4{g^2}{b}|{\bf w}|^2{\bf w}  \label{w1}\\
&+& \4{g {\rm Pe}_{\rm r}}{4 b}\left[5\nabla {\bf w}^2 - 10{\bf w}(\nabla \cdot {\bf w}) -6 ({\bf w}\cdot \nabla){\bf w}\right] \nonumber \\
&+&\Omega {\bf w}_\perp  +\4{{\rm Pe}_{\rm r}^2 \Omega}{4b}\nabla^2 {\bf w}_\perp - \4{g^2 \Omega}{2 b}|{\bf w}|^2 {\bf w}_\perp \nonumber \\ 
&-& \4{g {\rm Pe}_{\rm r} \Omega}{8 b}\left[3\nabla_\perp {\bf w}^2 - 6{\bf w} (\nabla_\perp \cdot {\bf w}) - 10({\bf w} \cdot \nabla_\perp) {\bf w} \right] \nonumber 
\ee
Here $b=2(4+\Omega^2)$, ${\bf w}^{(1)}_\perp=(-w_y^{(1)},w_x^{(1)})$, and $\nabla_\perp=(-\partial_y,\partial_x)$.
In the special case $\Omega=0$, when neglecting second order derivatives (\ref{rho1},\ref{w1}) are identical to the limiting case of a density-independent swim speed in \cite{Farrell2012sm}.

\subsection{Flocking in circle swimmers: enhanced flocking}
Eqs.~(\ref{rho1},\ref{w1}) have two uniform solutions representing the disordered uniform phase (D) 
$(\rho,{\bf w})=(\rho_0,{\bf 0})$, where $\rho_0$ is fixed by the initial conditions and conserved in the course of the dynamics (\ref{rho1}), and a uniform flock (F)
$(\rho,|{\bf w}|,\dot \phi)=(\rho_0,w_0,\omega)$ (where $\phi({\bf x},t)$ is defined via ${\bf w}/w=(\cos \phi,\sin \phi)$) which features long-range polar order in two-dimensions
\1 w_0 = \4{1}{g}\sqrt{(g \rho_0-2)(4+\Omega^2)} \label{w1pol}\2
and rotates with a frequency 
\1 \Omega_0 = \Omega \left(\4{3}{2}-\4{g \rho_0}{4}\right) \label{phipol}\2
Remarkably, following (\ref{w1pol}), rotations enhance the degree of polar order in the flocking phase (enhanced flocking), as discussed in more detail in the main text. 
Interactions in turn, lead to a slowdown of rotations of the flock which 
changes direction with the frequency of the underlying circle swimmers $\Omega_0=\Omega$ at the onset of flocking ($g \rho_0=2$) and slows down as more particles align (see (\ref{phipol})). 
\\Linearizing (\ref{rho1},\ref{w1}) (D) shows that the disordered phase gets unstable at $g\rho_0=2$, which is the ordinary flocking transition. Hence, 
independently of how strong rotations are, the emergence of long-range order 
solely depends on a competition of noise and alignment interactions. 
In other words: identical rotations of all swimmers are irrelevant for structure formation in the RVM in absence of polar order ($g \rho_0<2$).
This finding crucially changes as soon as polar order emerges, as we now demonstrate. 

\subsection{Pattern formation in circle swimmers: A linear stability analysis of the flocking phase}
To understand the onset of structure formation in the RVM, we now perform a linear stability analysis of the uniform flocking phase (F).
As we will see, in this phase, circular swimming of individual particles dramatically changes the phenomenology as compared to the standard Vicsek model and
creates a route to the formation of patterns whose length scale grows linearly with ${\rm Pe}_r$ and decreases with $\Omega$.

As usual, to test the stability of the flocking state we calculate whether a small perturbation on top of it 
grows or decays.
We therefore linearize (\ref{rho1},\ref{w1})
around (\ref{w1pol},\ref{phipol}), i.e. we use
$(\rho, {\bf w})=(\rho_0,{\bf w}_0)+(\rho',{\bf w}')$ 
with primes denoting fluctuations and transform the result to Fourier space. 
Generally, the rotation of the base state (F) produces time-dependent coefficients in some terms. 
In most cases, however, we will see, that the maximum growth rates of unstable modes in the RVM at a given orientation of ${\bf w}$ strongly exceed $\Omega$; e.g. by one decade in Fig.~\ref{fig1sm}, left).
Thus, the flock does not rotate much on the timescale where perturbations grow and drive the system out of the linear regime. 
Therefore, we perform our linear stability analysis at a given orientation of ${\bf w}$, 
leading to the following linearized equations of motion:
\ea
\left(\begin{matrix} \dot \rho' \\ \dot w'_x \\ \dot w'_y \end{matrix}\right) = 
\left(\begin{matrix}
0 & \I {\rm Pe} q_x & \I {\rm Pe} q_y \\
\4{g w_0}{2} + \I \4{{\rm Pe}}{2}q_x & (2-g \rho_0) + \I r \left(3q_x+\4{5\Omega}{2}q_y\right) -\4{{\rm Pe}^2 {\bf q}^2}{2b}& -\Omega_0 + \I r \left(5q_y-\4{3\Omega}{2}q_x\right) + \4{\Omega {\rm Pe}^2 {\bf q}^2}{4b}\\ 
\I \4{{\rm Pe}}{2}q_y & (1-\4{g\rho_0}{2})\Omega+\Omega_0 - \I r \left(5 q_y-\4{3\Omega}{2}q_x \right) -\4{\Omega {\rm Pe}^2 {\bf q}^2}{4b} &  \I r\left(3q_x + \4{5\Omega}{2}q_y\right) - \4{{\rm Pe}^2 {\bf q}^2}{2b}
\end{matrix}\right) 
\left(\begin{matrix}\rho' \\ w'_x \\ w'_y \end{matrix}\right)
\label{stabmat}
\ee
Here ${\bf q}=(q_x,q_y)$ is the wavevector, $r=\4{{\rm Pe} g w_0}{2b}=\4{{\rm Pe}}{4}\sqrt{\4{g\rho_0-2}{4+\Omega^2}}$ and $b=2(4+\Omega^2)$.

Despite its rather complicated appearance, (\ref{stabmat}) allows for a number of useful observations:
\\(i) The Peclet number ${\rm Pe}_r$ can be absorbed in the wavenumbers $q_x,q_y$ in (\ref{stabmat}). Thus, 
linear stability criteria ('phase transition lines') are independent of the Peclet number and therefore in particular independent of the self-propulsion velocity (as long as ${\rm Pe}_r \neq 0$).
\\(ii) For the same reason, the length scale of any pattern arising via a linear instability from the flocking solution will scale as $l^\ast \propto {\rm Pe}_r$.
Such a scaling can be observed for the microflock pattern as we confirm with particle based simulations in the main text.
\\(iii) $g$ and $\rho_0$ appear only together as $g \rho_0$ in (\ref{stabmat}). Thus, the linear stability (or nonequilibrium 'phase diagram') depends only on two dimensionless parameters: $g \rho_0$ and $\Omega$
and therefore, the two-dimensional plot of the phase diagram shown in the main text is representative for the complete parameter space of the RVM
(whose dynamics generally depends on 7 (4) parameters before (after) transforming to nondimensional units. 

We now proceed with a more formal analysis of (\ref{stabmat}).
The flocking phase is unstable if at least one of the eigenvalues of the matrix in 
(\ref{stabmat}) has a positive real part at some wavenumber ${\bf q}$. 
As the base state rotates slowly compared to the growth rate of fluctuations, we can analyse the stability of perturbations parallel ($q_y=0$) and perpendicular ($q_x=0$) to the polarization direction separately, 
as usual for nonrotating systems. 
Since the dispersion relation $\lambda(q_x,q_y)$ is a complicated high order polynomial in $q_x,q_y, \Omega$ and $g \rho_0$, we apply various 
approximations to roughly understand the onset structure formation. 
The resulting instability criteria are summarized in an instability or nonequilibrium phase diagram in Fig.~4 of the main text.

\subsubsection{Instabilities along polarization direction ($q_y=0$)}
We first analyse the response of the standard Vicsek model to small perturbations parallel to the flocking direction for the standard Vicsek model ($\Omega=0$).
Expanding the dispersion relation $\lambda(q_x)$ to second order in $q_x$ around $q_x=0$
unveils an oscillatory long wavelength instability for $2<g \rho<22/7$. This instability is often associated with the emergence of travelling bands in the Vicsek model
if the density is not too large. 

To see how rotations affect this instability, we now expand
$\lambda(q_x)$ to second order both in $q_x$ (around $q_x=0$) and in $g \rho_0$ (around the flocking threshold $g\rho=2$). 
As a result, we find that the same oscillatory long wavelength instability is always present in the RVM and hence robust against arbitrarily fast rotations. 
To see if this result also holds true further away from the flocking threshold, we now expand $\lambda(q_x)$
both in $\Omega$ and $q_x$ to second order around 0. 
As a first result, we find that any $\Omega>0$ destabilizes phase (F) even at zero wavenumber (${\bf q}=0$) if $g\rho_0>10/3$.
This suggests that the RVM allows for long-wavelength instabilities even at interaction to noise ratios where the standard Vicsek model is deep in the uniform flocking phase. 
(More generally, this result also follows by considering $\lambda({\bf q})$ for ${\bf q}=0$ without expanding in $\Omega$.)
The regime $22/7<g\rho_0<10/3$ is more involved: the same expansion in $\Omega,q_x$ shows that fast enough rotations can induce the 
long-wavelength instability also at moderate $g\rho_0$ values, where fast enough is quantified by\footnote{We note that the derivation of our 
continuum theory assumed that we are close to isotropy, so it is unclear if this regime really exists.}
\1 \Omega> 4 \sqrt{\4{(g\rho_0-2)(14-g\rho_0)^3}{g\rho_0 \left[49120-g\rho_0(12808 + 3g\rho_0(9g\rho_0-424) ) \right] -64944}}\2

Besides the long wavelength instability we also find a short wavelength instability in polarization direction.
However, a quantitative criterion for this instability is quite involved as different modes 
(branches of the dispersion relation) can cross each other and the instability 
is in most cases caused by high order terms in $q_x$.
A numerical analysis of this instability shows that it typically masked by a corresponding short wavelength 
instability perpendicular to the flocking direction (which often has a larger growth rate) which we discuss below.

\subsubsection{Instabilities perpendicular to the polarization direction ($q_x=0$)}
We now explore the response of the RVM against small perturbations perpendicular to the polarization direction. 

\begin{figure}
\begin{center}
\includegraphics[width=0.7\textwidth]{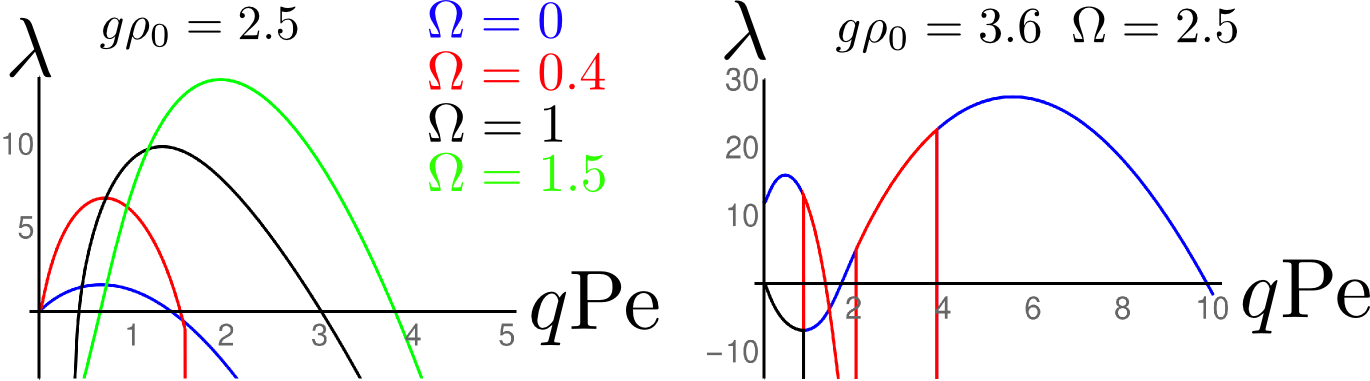}
\end{center}
\caption{\small Real part of the dispersion relation ${\rm Re}[\lambda(q_y)]$ (growth rates) of phase (F) perpendicular to the polarization direction: 
Left: Close to the flocking threshold, rotations can suppress the long wavelength instability perpendicular to the flocking direction and generate 
an oscillatory short wavelength instability ($\Omega=1.0; 1.5$) leading to microflock patterns. For slow enough rotations $\Omega=0.4$ the long wavelength instability of the Vicsek model survives but turns into an 
oscillatory instability contributing to the emergence of rotating macro-droplets (see Fig.~2 in the main text). 
Right: Further away from the flocking threshold ($g\rho_0=3.6$) rotations can lead to coexisting short and long wavelength instabilities which are separated by a 
band of stable wavenumbers (colors represent the different branches of the dispersion relation for fixed parameter values).} 
\label{fig1sm}
\end{figure}

\vspace{0.5cm}
\textbf{Long Wavelength Instability:}
We first consider the standard Vicsek model ($\Omega=0$) again. Expanding $\lambda(q_y)$ unveils a stationary long-wavelength instability 
perpendicular to the polarization direction, for $2<g\rho_0<82/21 \approx 3.9$.\footnote{
Remarkably, in a small parameter window, for $g\rho_0 \in (3.4,3.6)$ we also find an oscillatory short wavelength instability for the standard Vicsek model, which coexists with the 
long-wavelength instability and is separated from it by a gap of stable wavenumbers. While this suggests, in principle, the existence of some kind of dynamic non-coarsening pattern 
in the Vicsek model, such structures may not be observed in practice as the growth rate of the short wavelength structures is very small.}
(This perpendicular instability has not been discussed much in the literature; one exception is \cite{Gopinath2012sm} 
where a corresponding instability was analysed and discussed but in a more phenomenological model.)

For the RVM ($\Omega \neq 0$) we expand the relevant branch of the dispersion relation $\lambda(q_y)$ up to second order around $q_y=0$ and $g\rho_0=2$.
In presence of rotations, we find a corresponding perpendicular long wavelength instability if 
\1 \Omega<\sqrt{(8g\rho_0-16)/(7g\rho_0-5)} \label{perplg}\2
That is, rotations tend to suppress this long wavelength instability close to the flocking threshold; we visualize this in the phase diagram, Fig.~4, 
in the main text. Further away from the threshold, for $g\rho_0>10/3$, as mentioned above, any slow rotation generates a long-wavelength instability even at ${\bf q}=0$.
Remarkably, while the long wavelength instability perpendicular to the flocking direction is stationary for the 
standard Vicsek model it is oscillatory for the RVM and plays an important role 
for the emergence of the rotating droplets as we discuss in the main text.
\\To quantitatively compare the parameter domain where this instability exists with numerical simulations (see phase diagram, Fig.~4, in the main text)
we now generalize (\ref{perplg}), by expanding $\lambda(q_y)$ to third order in $g\rho_0$, which leads to
\1 \Omega<8 \sqrt{\4{(g\rho_0-2)(3g\rho_0-4)}{196 +g\rho_0 (69g\rho_0 -164)}}\2

\vspace{0.5cm}
\textbf{Microflock instability - Short wavelength modes:}
Most important to pattern formation in the RVM are short wavelength fluctuations perpendicular 
to the polarization direction. Identifying the branch of the dispersion relation which is most 
relevant for short wavelength instabilities and expanding it to second order around $g\rho_0=2$ 
and to first order around $q_y=0$, we find an oscillatory short 
wavelength instability if
\1 \Omega > \Omega_{\rm cr}=\sqrt{\4{4g \rho_0-8}{12-g\rho_0}} \label{mfinst}\2
This criterion holds true for $g\rho \gtrsim 2$ and leads to a complex $\Omega_{\rm cr}$ for $g\rho_0$ revealing 
that the corresponding instability only exists in presence of rotations.
For $\Omega>0$ however, the transversal short wavelength instability generically exists close to the flocking 
threshold and leads to pattern formation in the RVM. 
This instability creates microflocks with a self-limited size $l=2\pi/q_{\rm m}$ with 
$q_{\rm m}$ being the long wavelength of the associated instability band, which reads
\1 
q_y \approx \4{4\Omega^2}{{\rm Pe}_r} \4{|4(2-g\rho_0)+\Omega^2(12-g\rho_0)|}{\sqrt{(g\rho_0-2)(4+\Omega^2)}} \label{qyscal}
\2
The microflock length scale $l$ increases 
linearly with the Peclet number as expected from our more general considerations above. 
It also increases with $g\rho_0$ and decreases with $\Omega$ (the latter holding true at least not too close to onset of this instability). 
While the scaling law (\ref{qyscal}) should be precise only close to the 
$g\rho_0=2$-flocking onset, we find that the qualitative scaling applies more generally as a numerical analysis of the dispersion relation shows. 
In the main text, we confirm these scaling predictions using particle based simulations. 
\\To quantitatively compare our prediction for the onset of pattern formation in the RVM
with numerical simulations (main text), we now slightly generalize
(\ref{mfinst}), by expanding $\lambda(q_y)$ to third order in $g\rho_0$ which allows for a feasible result:
\1 \Omega > 2 \sqrt{\4{160}{164+g\rho_0(12-7g\rho_0)}-1} \label{mfinst2} \2
\\For completeness, we finally account also for terms of order $q_y^2$; here, we expend $\lambda(q_y)$) both in $q_y$ and to second order in $g\rho_0$. Among more complicated expressions 
resembling (\ref{mfinst2}) 
this expansion shows that the short wavelength instability perpendicular to the flocking direction is generally present if $\Omega > 2\sqrt{2/7}$.
\\We finally note, that long wavelength instabilities both in (and perpendicular to the) polarization direction 
typically (partly) coexist with the short wavelength instability in the RVM (compare Fig.~\ref{fig1sm}).  
This suggests that a given parameter allows for coexisting routes both towards phase separation and pattern formation. In this regime, the initial conditions decide which type of structure emerges (hysteresis) as we confirm 
with simulations (Movie 3). 

We summarize the instabilities perpendicular to the flocking direction in a nonequilibrium phase diagram in the main text, where we compare them with simulations.

\section{Brownian Dynamics simulations of Circle Swimmers}

Here we provide some details about the numerical simulations of the RVM. In particular, we specify the specific parameters used and the different measurements done in order to obtain the results presented in the main text.

\subsection{Numerical details and method}

We solved numerically the Langevin equations (1) and (2) in the main text using the Euler integration method with a time step $\Delta t=0.1$. We simulated system of particles moving in two dimensions in a $L\times L$ squared box with periodic boundary conditions.  For all the simulations presented in this work the average density and the rotational diffusion coefficient are fixed to $\rho_0=N/L^2=20$ and $D_r=0.5$.
To account for finite-size effects, we run simulations with $N=2000$, $8000$ and $32000$. We vary the coupling strength from $g=0$ to $g=0.4$, the self-propulsion velocity from $v=0.1$ to $v=1.5$ and the rotation frequency from $\omega=0$ to $\omega=2$ ($\Omega=4$ in adimensional units).  In order to reach the steady state we let the system evolve over more than $10^6$ time steps. We took special care in making sure that the system has reached the stationary state by looking at space-time correlation functions. We found that the formation of the patterns described in the main text is a slow process and one needs to let the system relax over time scales of this order of magnitude to be able to make any reliable measurement.  

\begin{figure}
\begin{center}
\includegraphics[scale=0.7,angle=0]{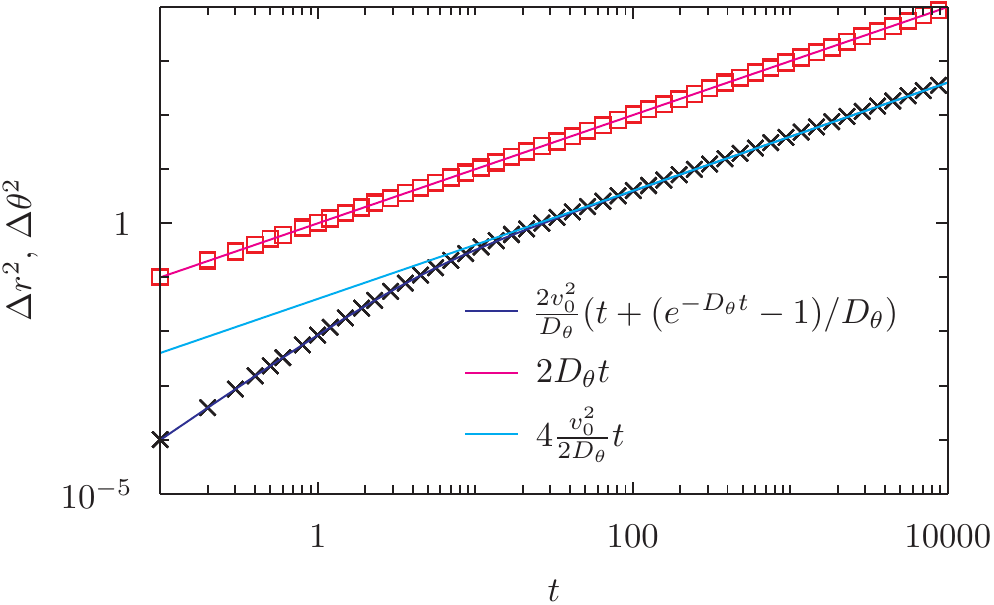}
\includegraphics[scale=0.7,angle=0]{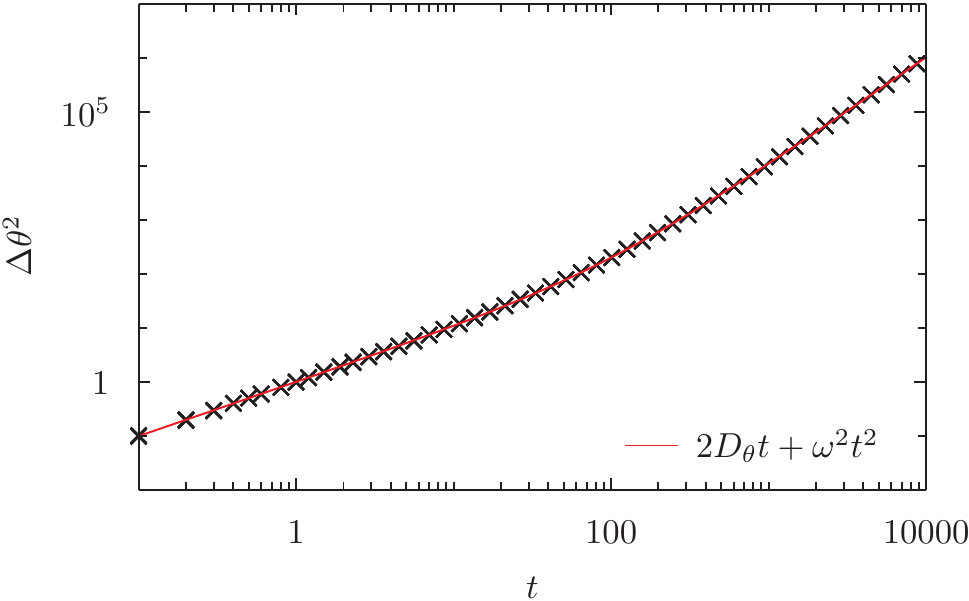}
\end{center}
\caption{Single particle mean-squared displacements. Left: Translational $\Delta r^2$ and angular $\Delta \theta^2$ mean-squared displacement (in black and red dots respectively) in the absence of active rotations in the dilute limit ($g=0$).  Right: Angular  mean-squared displacement in the dilute limit for a rotation frequency $\omega=0.1$. Continuous lines correspond to the analytical results eqs. (15-17).} 
\label{fig:MSD}
\end{figure}

In order to provide a simple check  of our simulation scheme we compare the mean-squared displacement of a single self-propelled particle obtained numerically with the analytical solution of the Langevin equations.  In the non-interacting limit, the position variables should perform a persistent random walk characterized by $v$ and $D_r$. The motion of the particles is diffusive at long time scales compared to the persistent time $\tau=1/D_{\theta}$. The mean square displacement can be computed analytically and gives,
\begin{equation}\label{eq:MSDdilute}
\langle \Delta \boldsymbol{r}^2(t)\rangle=\langle(\boldsymbol{r}(t)-\boldsymbol{r}(0))^2 \rangle=-\frac{v^2}{D_{\theta}}\left(\frac{2-e^{2D_rt} - e^{-D_rt}}{D_r}\right) \ , 
\end{equation}
which in the high persistence regime it can be approximated by
\begin{equation}\label{eq:MSDdilute}
\langle \Delta \boldsymbol{r}^2(t)\rangle=4\frac{v^2}{2D_r}\left( t+ \frac{1}{D_r}(e^{-D_rt}-1)\right) \ .
\end{equation}
In the dilute limit, the angular variables should verify
\begin{equation}
 \Delta \theta^2=\langle(\theta(t)-\theta(0))^2 \rangle=\omega^2 t^2+2D_rt \, .
\end{equation}
As shown in Fig.~\ref{fig:MSD}, our simulation method reproduces these results accurately. 

\subsection{Flocking transition}
\begin{figure}
\begin{center}
\includegraphics[scale=0.7,angle=0]{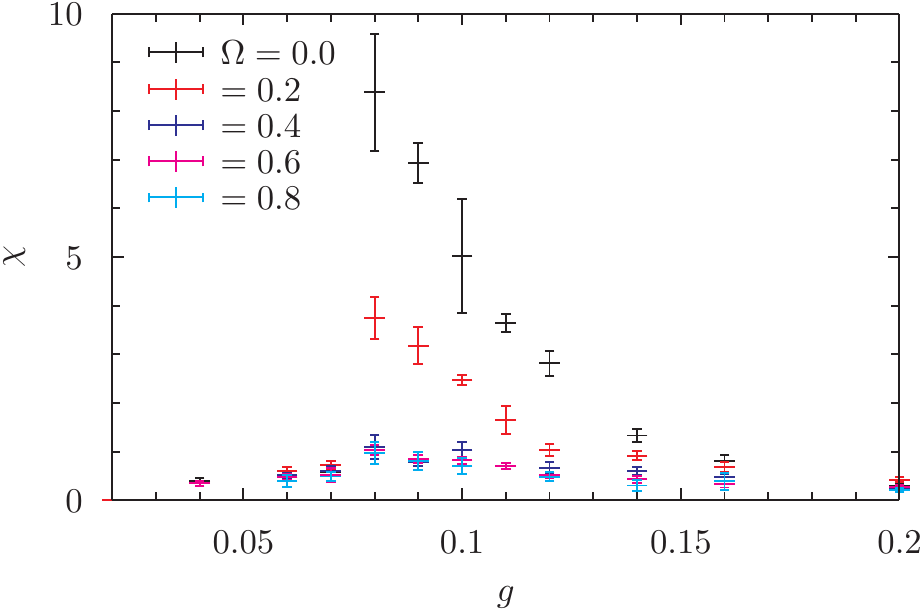}
\end{center}
\caption{Susceptibility as a function of $g$ for several values of $\Omega$.  The peak of $\chi$ indicating a phase transition is at $g\approx 0.08$ independently of $\Omega$. The value predicted by the hydrodynamic theory is $g_f=0.1$, slightly above the numerical measurement. } 
\label{fig:chi}
\end{figure}

We focus first on the emergence of spontaneous polar order  as $g$ increases. We introduce the order parameter
\begin{equation}
P=\langle||\boldsymbol{p}||\rangle \ ,\, \  \boldsymbol{p}=N^{-1}\sum_i \boldsymbol{n}_i \, , 
\end{equation} 
and its  associated susceptibility
\begin{equation}
\chi=N\left[\langle\boldsymbol{p}^2\rangle-\langle\boldsymbol{p}\rangle^2\right] \, .
\end{equation}
The order parameter as a function of the coupling $g$ obtained for a system of $N=2000$ circle swimmers is shown in Fig.~3 in the main text. We show here in Fig. \ref{fig:chi} the corresponding susceptibility data. We identify the flocking transition with the maximum of the susceptibility. The phase boundary obtained in such a way is reported by black symbols in the phase diagram in the main text. 

 As it has been argued for the standard Vicsek model, finite-size effects are particularly relevant to determine the nature of the flocking transition \cite{VicsekRevsm}. The patterns, like traveling bands,  emerging in these systems can only be obtained in simulations of large enough systems. We did not attempt to provide here a full analysis of the flocking transition in this model.  This would require a precise finite-size scaling analysis. 
 As shown in Fig. \ref{fig:chi}, the amplitude of the order parameter fluctuations decreases with $\Omega$. It might indicate that rotations change the nature of the flocking transition. This is however a speculation and we postpone this issue to a future work. 
 
However, we systematically increase the size of our system in order to identify different patterns that are out of reach using small systems, since they are characterized by a length scale that might be of the order of the system size. Simulations of different system sizes also allows us to test the robustness of the results presented. Even larger systems than the ones investigated in this work would be needed in order to analyze the patterns at even higher couplings. The patterns are expected to grow with $g$ and the different instability mechanisms described above might lead to different patterns that can not be  properly identified with the simulations presented here.

\subsection{Microflocks}

As discussed in the main text, for fast enough rotations, we observe a change of morphology in the system. In practice,  the phase boundary between the phase separated region and the microflock phase divides states with a single macroscopic cluster from states with several smaller ones. In order to make a quantitative estimation of this phase boundary that allows comparison with the linear instability analysis of the hydrodynamic equations, we compute the cluster size distribution $\mathcal{P}_m$. We define a cluster as a connected set of particles distant of less that $1/3$ (in units of $R_{\theta}$). 

\begin{figure}
\begin{center}
\includegraphics[scale=0.65,angle=0]{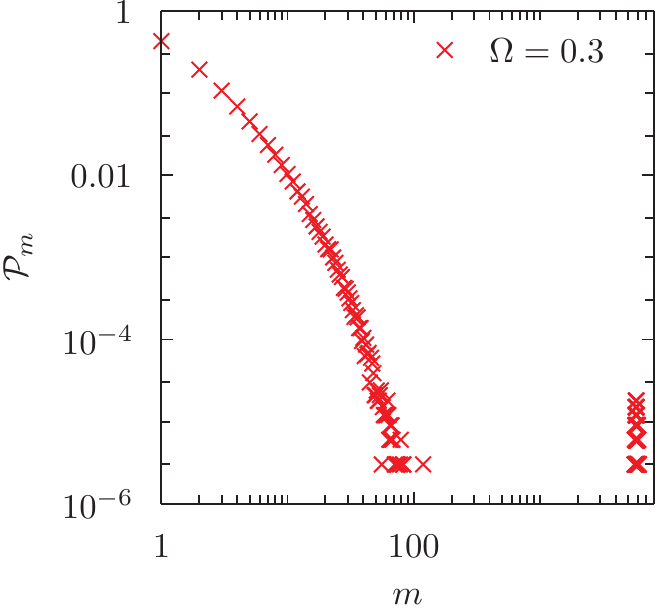}\includegraphics[scale=0.65,angle=0]{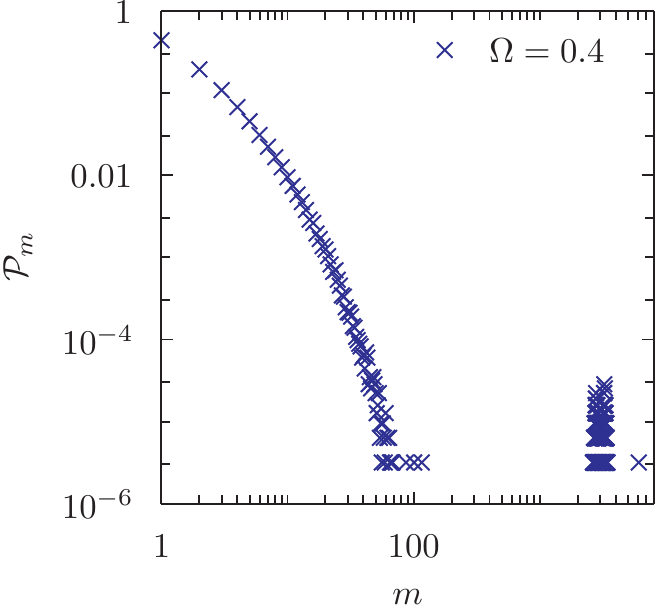}\includegraphics[scale=0.65,angle=0]{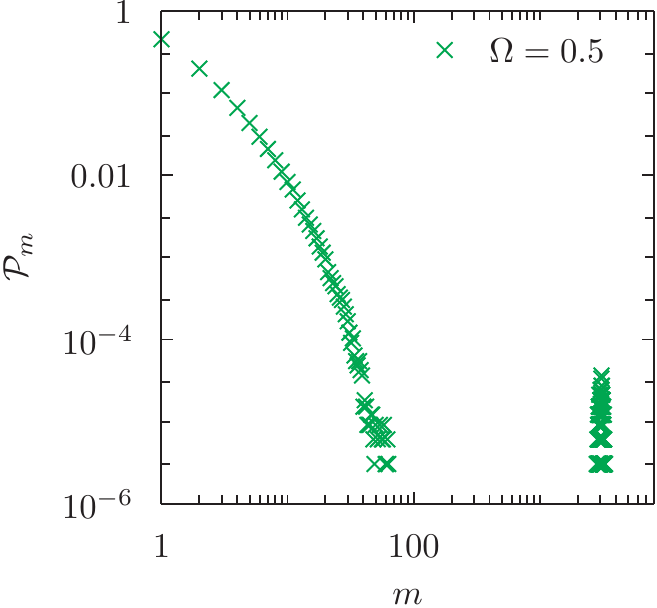}\includegraphics[scale=0.65,angle=0]{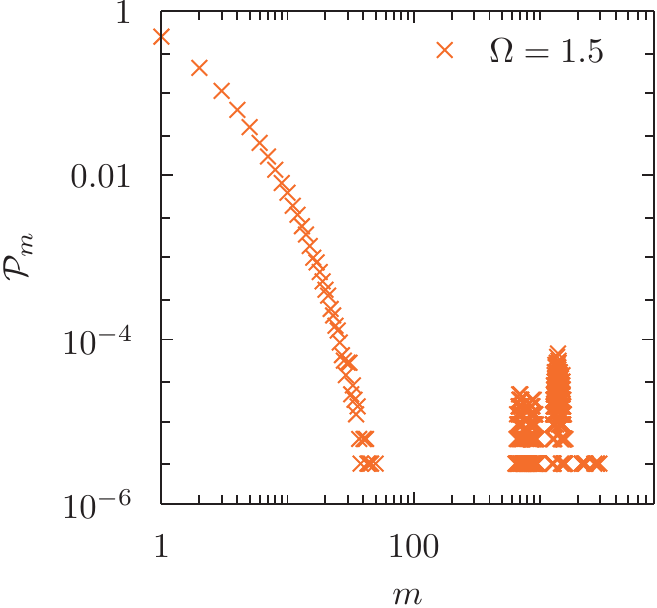}
\\
\hspace{1.1cm}\includegraphics[scale=0.2,angle=-90]{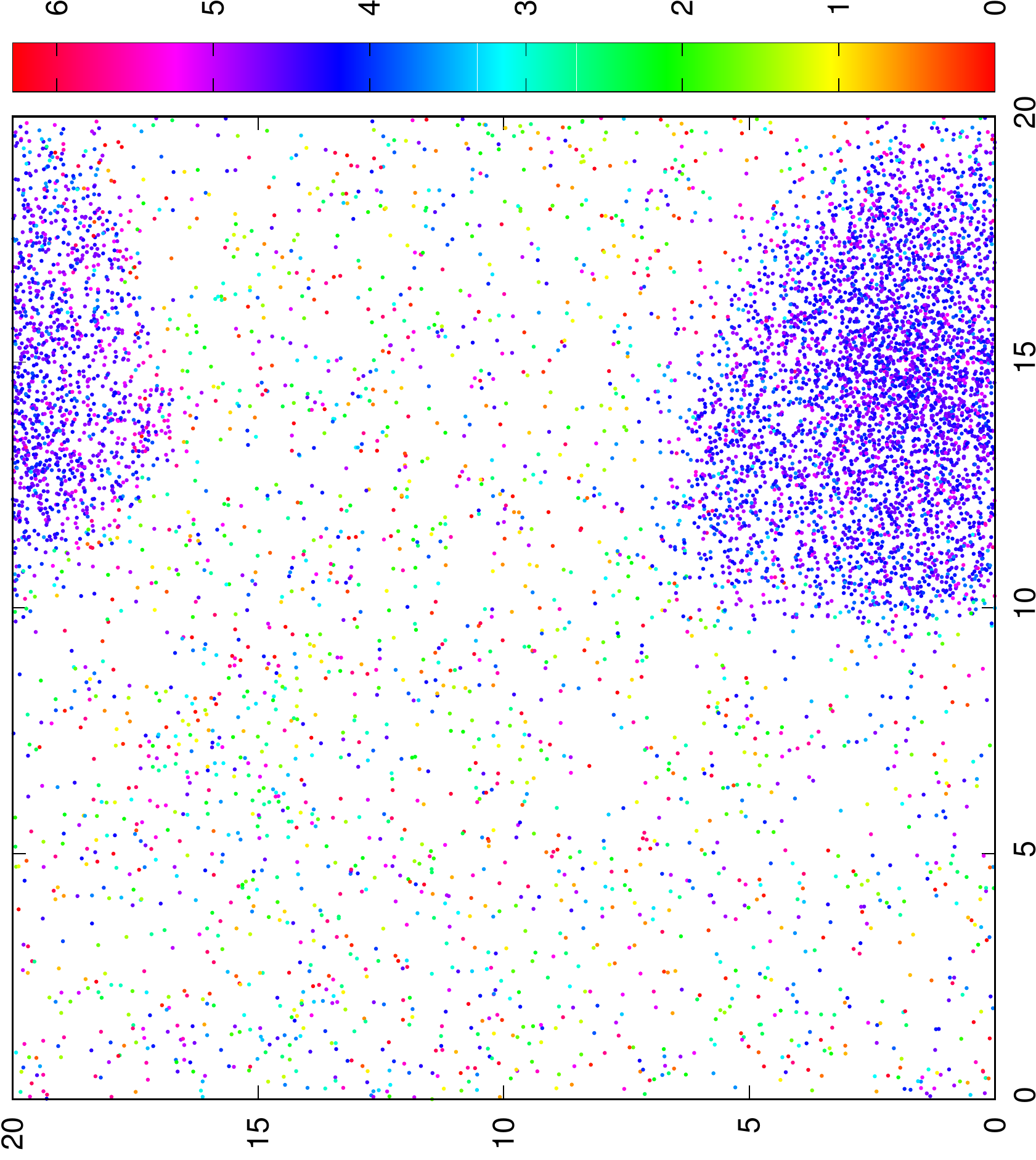} \hspace{0.3cm}
\includegraphics[scale=0.2,angle=-90]{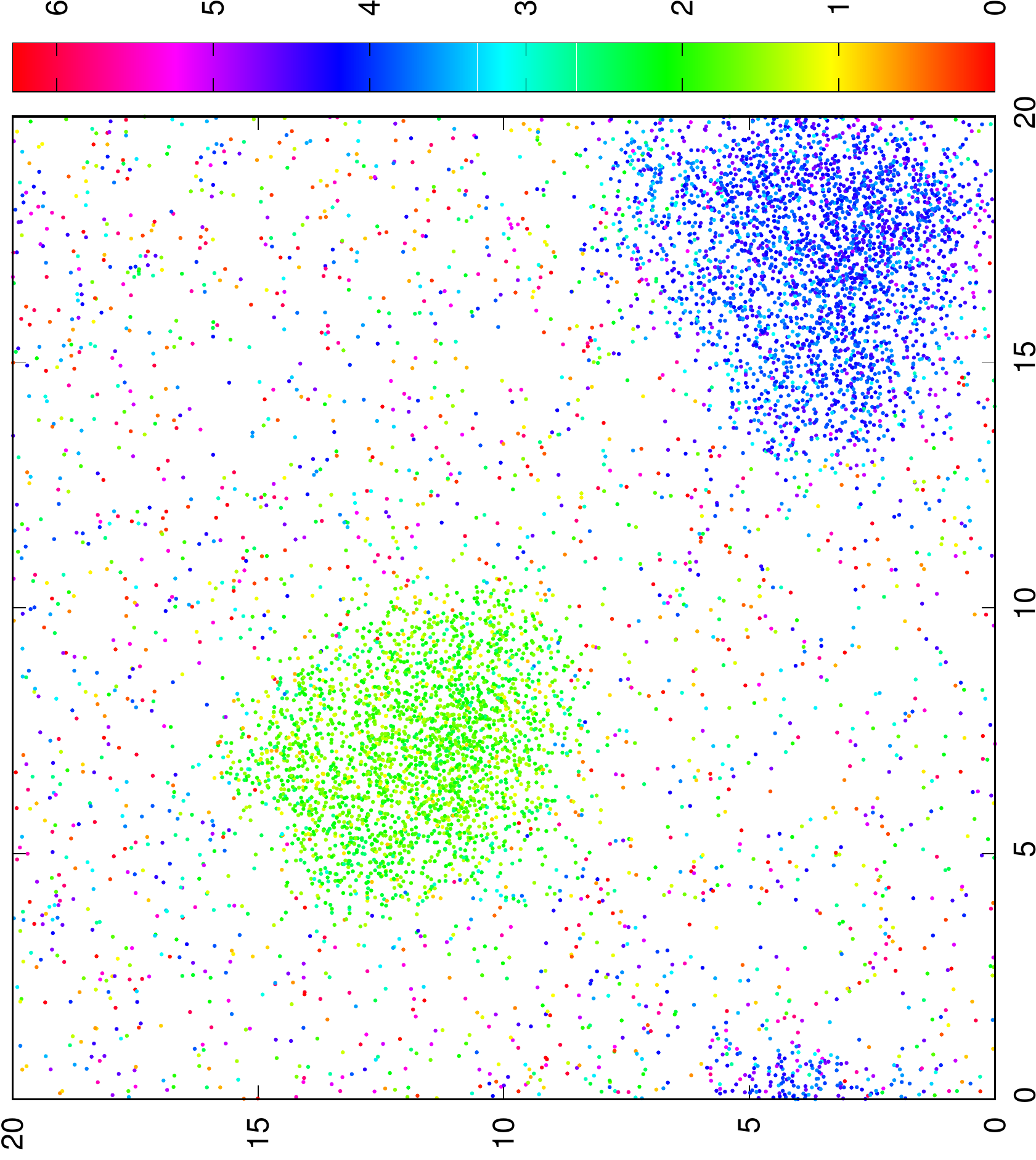} \hspace{0.3cm}
\includegraphics[scale=0.2,angle=-90]{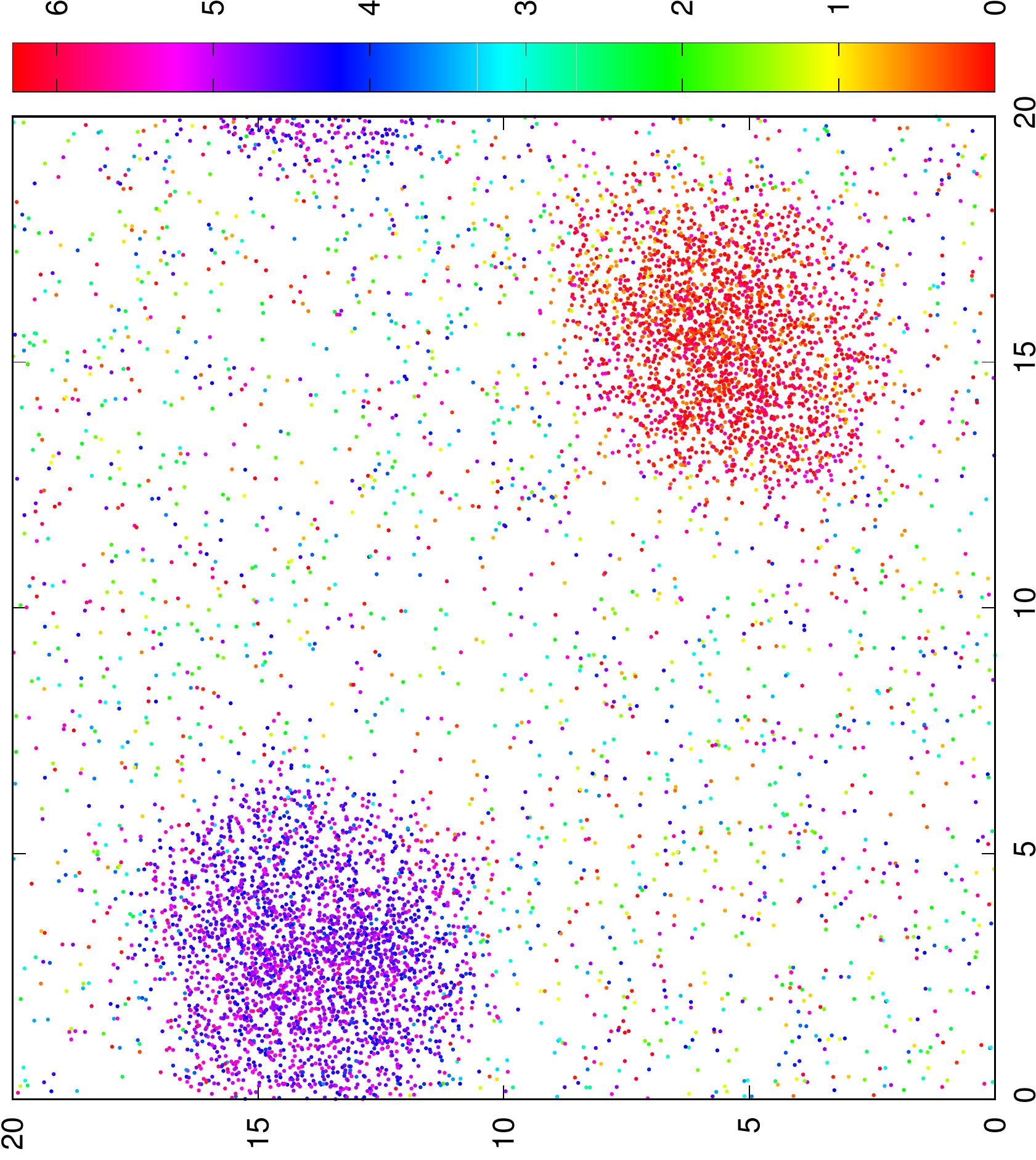} \hspace{0.3cm}
\includegraphics[scale=0.2,angle=-90]{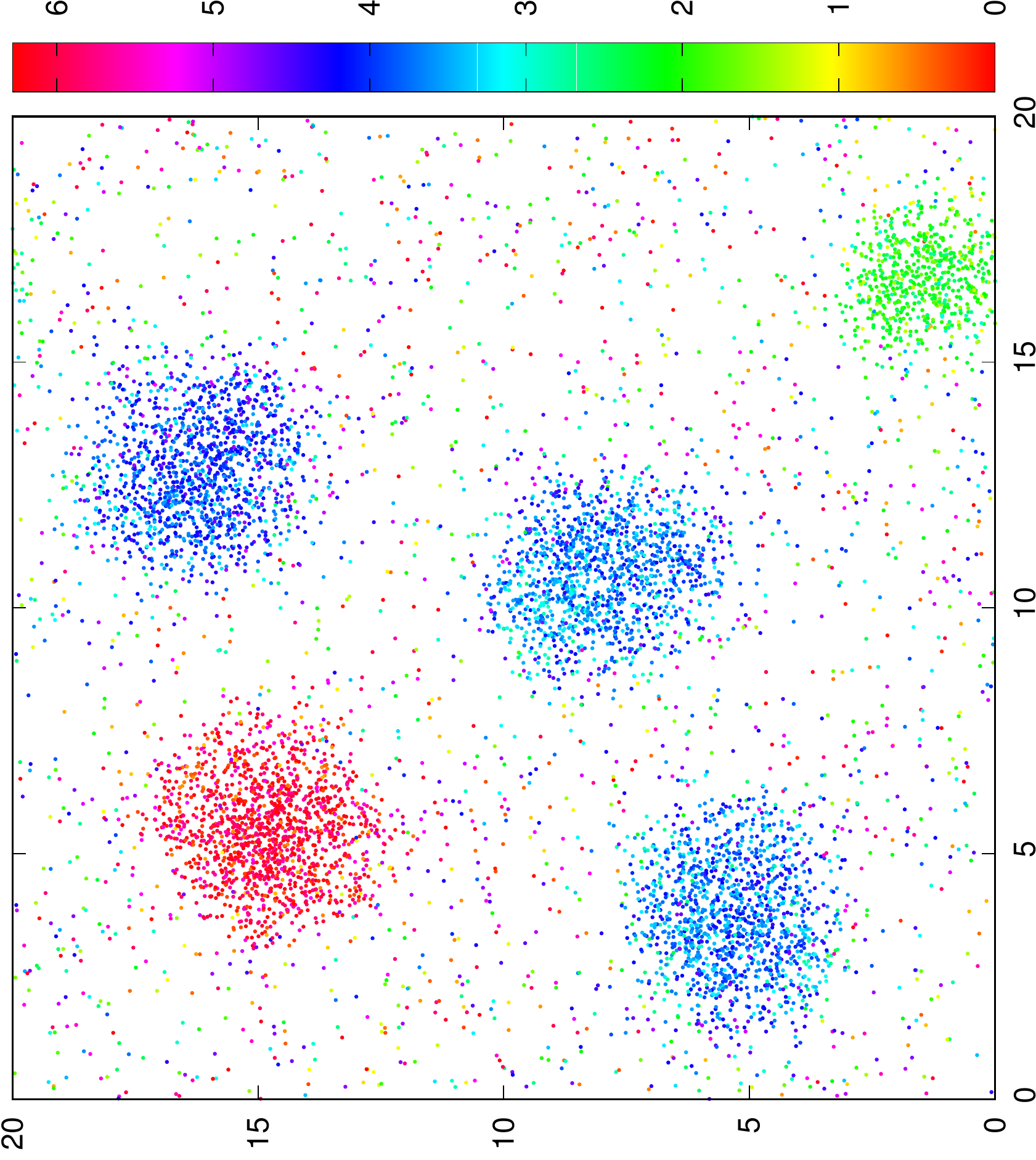}

\end{center}
\caption{Top: Cluster size distribution for $N=8000$, $g=0.11$, $v=0.1$ and several frequencies shown in the key. 
For  $\Omega=0.3$ a macroscopic cluster of size comparable with the system size appears. As we increase $\Omega$ the location of the peak(s) moves to lower system sizes, indicating the presence of smaller clusters.  The snapshots shown below confirm this picture.  We identify the phase boundary at $\Omega=0.4\pm0.1$.
Bottom: Snapshots of the steady state configuration corresponding to the distributions shown on top.
} 
\label{fig:CDF}
\end{figure}

The results for $g=0.11$ are shown in Fig. \ref{fig:CDF}. In the phase separated region, the distribution of clusters is characterized by the  coexistence between an exponential distribution of small clusters and a peak at cluster sizes of the order of the system size. In the presence of faster rotations, smaller clusters of a tunable finite size appear, which in the cluster size distribution translates into the presence of a peak at smaller values of $m$ as compared to the phase separated state. This change of behaviour in the distribution allows us to estimate the phase boundary between both phases, as reported in the phase diagram in the main text.      

\subsection{Movies}
For all the movies, the color code is the same as for Fig. 2 in the main text. 

\begin{itemize}
\item Movie 1: Evolution of a system made of $N=32000$ particles from an initial homogenous disordered state towards a phase separated state with $\Omega=0.2$ and $g=0.14$. Available at: https://drive.google.com/file/d/0B5Gy3WsV8841RlpqS3huRXNzOW8/view
\item Movie 2: Evolution of a system made of $N=32000$ particles from an initial homogenous disordered state towards a microflock state with $\Omega=3$ and $g=0.14$.  Available at: https://drive.google.com/file/d/0B5Gy3WsV8841TXFnU1hXNmFxZkk/view 
\item Movie 3: Evolution of a system made of $N=32000$ particles from an initial inhomogenous  state in the phase separated region (previously prepared with $\Omega=0.2$ and $g=0.14$) for which faster rotations $\Omega=3$ are turned on at $t=0$.  Available at: https://drive.google.com/file/d/0B5Gy3WsV8841WnJyODZOWlEyYU0/view
\end{itemize}

\bibliographystyle{apsrev}
\bibliography{./literature}

\end{document}